\documentclass[aps,prd,twocolumn,amsmath,amssymb,nofootinbib,showpacs]{revtex4}
\usepackage{graphicx,slashed,cancel,bbm}
\newcommand{\be}{\begin{equation}}
\newcommand{\ee}{\end{equation}}
\newcommand{\bea}{\begin{eqnarray}}
\newcommand{\eea}{\end{eqnarray}}

\newcommand{\sfrac}[2]{{\textstyle\frac{#1}{#2}}}
\newcommand{\D}{\mathrm{d}}
\newcommand{\E}{\mathrm{e}}
\newcommand{\I}{\mathrm{i}}
\newcommand{\Det}{\mathrm{Det}}
\newcommand{\Tr}{\mathrm{Tr}}
\newcommand{\Lag}{\mathcal{L}}
\newcommand{\m}{\mu}
\newcommand{\hta}{\check\tau}

\newcommand{\dy}{z}
\begin{document}
\title{The fifth dimension in the worldline formalism, holography, and the Wilson flow}
\author{Dennis D.~Dietrich}
\affiliation{Arnold Sommerfeld Center, Ludwig-Maximilians-Universit\"at, M\"unchen, Germany}
\affiliation{Institut f\"ur Theoretische Physik, Goethe-Universit\"at, Frankfurt am Main, Germany}
\begin{abstract}
We develop a model for hadrons in the framework of the worldline formalism.
While being based wholly in four-dimensional quantum field theory, it shares many features with holographic approaches: Already by the use of the worldline formalism the approach appears intrinsically quantum mechanical. As auxiliary fifth dimension Schwinger's proper time is grouped with the physical four spacetime dimensions into an AdS$_5$ geometry, which is warped due to conformal-symmetry breaking effects. Hidden local symmetry is emergent. The four-dimensional sources are extended to five-dimensional fields by a Wilson flow (gradient flow). A variational principle for the flow reproduces exactly the corresponding holographic computation. The worldline approach also yields the higher-dimensional description in the non-relativistic case.
\end{abstract}
\pacs{
11.25.Tq 
12.40.Yx 
}
\maketitle

\section{Introduction}

Strong interactions lead to a wealth of phenomena, but they also frequently overwhelm our computational abilities. Holographic approaches offer analytic insight and have been used intensely in quantum chromodynamics (QCD) \cite{Erlich:2005qh,Karch:2006pv,Polchinski:2000uf}, in corresponding extensions of the Standard Model \cite{Hong:2006si,Dietrich:2008ni}, and solid state physics \cite{Sachdev:2011wg}. Holography is based on the conjectured AdS/CFT correspondence \cite{Maldacena:1997re} and its extensions. All known examples of this correspondence, however, hold for theories with a set of symmetries that are not realised in nature (first of all supersymmetry) and which, already for this reason, have a different particle content. Therefore, usually deformed bottom-up AdS/QCD descriptions are considered, which describe the QCD hadron spectrum surprisingly well \cite{Karch:2006pv,Da Rold:2005zs}. They do, however, lack a derivation from first principles. Consequently, it is very important to understand for which reasons and under which circumstances these models represent an acceptable approximation.

In particular, the original correspondences are between field theories on four-dimensional Minkowski space and gravity/string theories in higher dimensional spaces. The gravity/string theory side is seen as an effective description of nature and not interpreted as reality (just as the scalar field in the Ginzburg-Landau model \cite{Ginzburg:1950sr} is not regarded as fundamental). What is customarily used in bottom-up approaches are generalised effective field theories on a warped AdS$_5$ space with a  restricted set of operators, which is admissible due to an efficient decoupling of higher operators \cite{Fitzpatrick:2013twa}. In this form, they map four-dimensional quantum field theoretical computations onto five-dimensional quantum mechanical ones. Hence, it is crucial to comprehend how this five-dimensional description comes to pass without recourse to gravity/string theory. Without the warping the extended spacetime possesses conformal symmetry in addition to Lorentz invariance, and conformal symmetry is used as lowest order approximation. Classical massless QCD, for example, is conformally invariant, and approximately so (at least) in the ultraviolet (asymptotic freedom). 
In quasiconformal (e.g., technicolour) theories this is an even better approximation \cite{walk}.
In order to take into account phenomena that violate conformal symmetry such as confinement, conformal symmetry is broken by introducing a scale, e.g., by warping or truncating the spacetime. Beyond capturing the isometries the physical significance of the extra dimension needs to be clarified. In \cite{deTeramond:2008ht}, for example, the extra coordinate is identified with $\zeta^2=x(1-x)\mathbf{b}_\perp^2$, where $x$ stands for the light-front momentum fraction of one of the constituents of the meson and $\mathbf{b}_\perp$ for the transverse separation of the constituents.

The present investigation is based on the worldline formalism \cite{Strassler:1992zr} for quantum field theory and continues the research initiated in \cite{Dietrich:2013kza}. The organising principle behind the derivation arises from the observation that additional gluons in bound-state wave functions appear to be very costly and thus rare. Take large-angle hadron-hadron scattering for example. There the exchange and/or annihilation of quarks \cite{White:1994tj} dominates the scattering process. (See also \cite{Gunion:1972qi} vs \cite{Landshoff:1974ew}.) The leading diagrams are thus of the form depicted in Fig.~\ref{fig}. Furthermore, according to the Okubo-Zweig-Iizuka (OZI) rule \cite{Okubo:1963fa} reactions that proceed exclusively via the exchange of gluons, i.e., whose Feynman diagrams come apart when all gluon lines are removed, are known to be suppressed relative to those which keep together.
After all, hadronic bound states can be categorised based on their valence quark content, and, what is more, we do not observe a wealth of multiquark states or hybrids or glueballs.

The following section motivates the steps undertaken thereafter by a comparison of the worldline formalism in Sect.~\ref{sec:wlf} with the AdS/QCD approach in Sect.~\ref{sec:adsqcd}. Based on our observations up to that point, we modify the worldline formalism with the aim to obtain a description for hadrons on the worldline, in Sect.~\ref{sec:hotwl}. Sect.~\ref{sec:twobody} relates this new approach to one with two-body interactions. Sect.~\ref{sec:wilson} describes how the extension of four-dimensional sources to five-dimensional fields---a key ingredient of holographic descriptions---arises as Wilson flow (gradient flow) in the worldline approach and Sect.~\ref{sec:repro} how the known AdS/QCD result arises in this context from a variational principle. Sect.~\ref{sec:instanton} relates the hadrons on the worldline to the worldline instanton picture and the Gutzwiller trace formula. Sect.~\ref{sec:summary} summarises our findings.

\section{Motivation\label{sec:motivation}}

\subsection{Worldline formalism\label{sec:wlf}}

\begin{figure}[h]
\centerline{%
\includegraphics[width=\columnwidth]{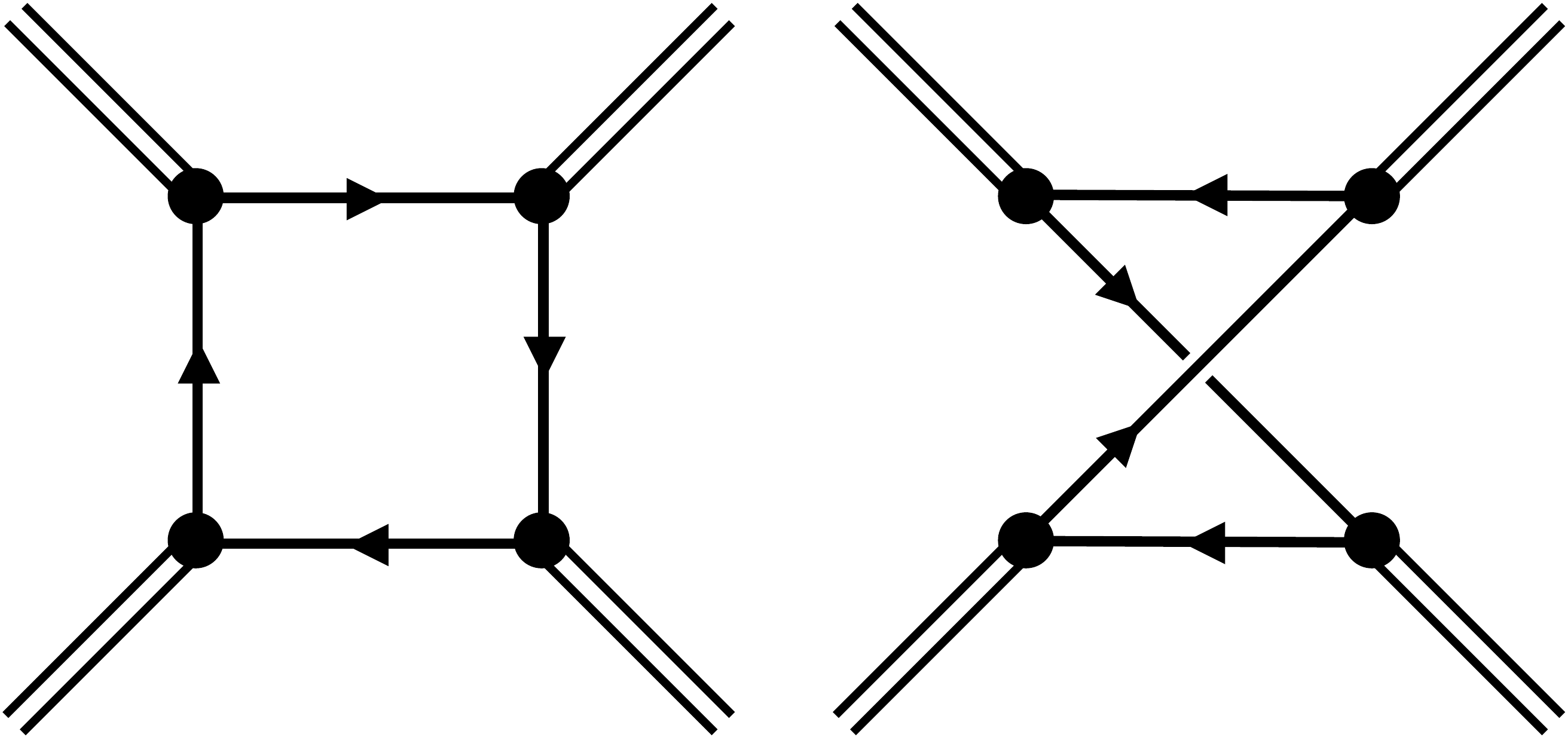}}
\caption{Some dominant diagrams. Double lines are for hadrons, single lines for quarks (taken from \cite{Dietrich:2013kza}).}
\label{fig}
\end{figure}
In order to demonstrate how the extradimensional structure materialises, in particular precisely the correct powers of the extradimensional variable necessary to have an AdS$_5$ structure, we provide a detailed derivation of the worldline formalism. The dominant diagrams, i.e., those without additional gluons, are contained in the effective action
\begin{align}
\ln Z
&=
\ln \int[\D\bar\psi][\D\psi]\E^{\I\int\D^4x\bar\psi(\I\slashed D-m)\psi}
=\\&=
\frac{1}{2}\ln\Det(\slashed D^2+m^2)
=\\&=
\frac{1}{2}\Tr\ln(\slashed D^2+m^2) .
\label{eq:genfun}
\end{align}
For the sake of simplicity we consider here merely a single flavour of mass $m$ and a vector source $V^\mu$, which we include in the `covariant derivative' $D^\mu=\partial^\mu-\I V^\mu$. We can add different sources and/or flavours straightforwardly. In order to avoid, for the time being, an additional finite dimensional trace or additional functional integration we regard scalar quarks, i.e., 
\be
w=\ln z = -\frac{1}{2}\mathrm{Tr}\ln(D^2+m^2).
\label{eq:scalar}
\ee
We shall comment on the case of elementary fermions below. Next, we represent the logarithm by an integral,
\be
w = \frac{1}{2}\mathrm{Tr}\int_{\varepsilon>0}^\infty\frac{\D T}{T}\E^{-T(-D^2+m^2)},
\label{eq:ln}
\ee
where we have rotated to Euclidean space. We must include a regularisation at the lower integration bound, which can be dropped once $w$ is normalised by dividing by another effective action, e.g., the source free one. In preparation for carrying out the trace, we time slice the exponential and approximate each factor up to linear order,
\begin{align}
w 
&= 
\frac{1}{2}\int_{\varepsilon>0}^\infty\frac{\D T}{T}\E^{-m^2T}\:\mathrm{Tr}\lim_{N\rightarrow\infty}\prod_{n=1}^N\E^{-\frac{T}{N}(-D^2)}
=\\&=
\frac{1}{2}\int_{\varepsilon>0}^\infty\frac{\D T}{T}\E^{-m^2T}\:\mathrm{Tr}\lim_{N\rightarrow\infty}\prod_{n=1}^N[1-\Delta\tau(-D^2)] ,
\end{align}
where $\Delta\tau=\frac{T}{N}$. Afterwards, we introduce one complete set of eigenstates of the position operator $\hat x$ and one of the momentum operator $\hat p$ after each slice, make use of their eigenvalue equations, replace the scalar products of the state vectors by Fourier phases, and undo the linear approximation of the exponential,
\begin{align}
&\prod_n[1+\Delta\tau D^2(\hat x,\hat p)]
=\\=
{}&\prod_n\int\frac{\D^4p_n}{(2\pi)^4}[1+\Delta\tau D^2(\hat x,\hat p)]|p_n\rangle\langle p_n|
=\\=
{}&\prod_n\int\frac{\D^4p_n}{(2\pi)^4}[1+\Delta\tau D^2(\hat x,p_n)]|p_n\rangle\langle p_n|
=\\=
{}&\prod_n\int\D^4x_n\frac{\D^4p_n}{(2\pi)^4}[1+\Delta\tau D^2(\hat x,p_n)]|x_n\rangle\langle x_n|p_n\rangle\langle p_n|
\nonumber=\\=\nonumber
{}&\prod_n\int\D^4x_n\frac{\D^4p_n}{(2\pi)^4}[1+\Delta\tau D^2(x_n,p_n)]|x_n\rangle\langle x_n|p_n\rangle\langle p_n|
=\\=
{}&\int\D^4x_{N+1}\prod_n\int\D^4x_n\frac{\D^4p_n}{(2\pi)^4}
\times\nonumber\\&\times
|x_1\rangle\E^{\Delta\tau D^2(x_n,p_n)+\I p_n\cdot(x_n-x_{n+1})}\langle x_{N+1}|,
\end{align}
where in the last step we have put in one more complete set of position eigenstates. Taking the trace using position eigenstates yields
\begin{align}
&\mathrm{Tr}\prod_n\E^{-\Delta\tau(-D^2)}
=\\=
{}&\int\D^4x\:\D^4x_{N+1}\prod_n\int\D^4x_n\frac{\D^4p_n}{(2\pi)^4}
\times\\&\times\nonumber
\langle x|x_1\rangle\E^{\Delta\tau D^2(x_n,p_n)-\I p_n\cdot(x_n-x_{n+1})]}\langle x_{N+1}|x\rangle
=\\=
{}&\prod_n\int_\mathrm{P}\D^4x_n\int\frac{\D^4p_n}{(2\pi)^4}
\E^{\Delta\tau[D^2(x_n,p_n)-\I p_n\cdot\frac{x_n-x_{n+1}}{\Delta\tau}]},
\end{align}
with the periodicity condition $x_{N+1}=x_1$ marked by `P,' and where we have factored out $\Delta\tau$ in the exponent. Performing the limit $N\rightarrow\infty$ leads to a path integral,
\begin{align}
&\lim_{N\rightarrow\infty}\mathrm{Tr}\prod_{n=1}^N\E^{-\Delta\tau(-D^2)}
=\\=
{}&\int_\mathrm{P}[\D x]\int[\D p]\:
\E^{\int_0^T\D\tau\{D^2[x(\tau),p(\tau)]+\I p(\tau)\cdot\dot x(\tau)\}}
=\\=
{}&\int_\mathrm{P}[\D x]\int[\D p]\:
\E^{\int_0^T\D\tau[-(p+V)^2+\I p\cdot\dot x]}
=\\=
{}&\int_\mathrm{P}[\D x]\int[\D p]\:
\E^{\int_0^T\D\tau[-\{p-\frac{\I}{2}\dot x+V\}^2-\frac{\dot x^2}{4}-\I p\cdot V]}
=\\=
{}&\frac{\pi^2}{(2\pi)^4T^2}\mathcal{N}\int_\mathrm{P}[\D x]\:
\E^{-\int_0^T\D\tau(\frac{\dot x^2}{4}+\I p\cdot V)}.
\label{eq:pathint}
\end{align}
In the last steps, we have integrated out the momentum field. The normalisation $\mathcal{N}$ cancels the free position path integral, $\mathcal{N}\times\int[dx]\E^{-\int_0^Td\tau\frac{\dot x^2}{4}}=1$. The other prefactor $\frac{\pi^2}{(2\pi)^4T^2}$ arises since there was one more momentum integration than position integrations. Below we have to invert the kinetic operator. Therefore, we must split off the zero mode arising from translational invariance, $x^\mu=y^\mu+x_0^\mu$, where $x_0^\mu$ must be constant along the path, $x_0^\mu\neq x_0^\mu(\tau)$. We can choose $x^\mu_0$, e.g., as `centre of mass' $\int_0^T\D\tau\,y^\mu=0$ or starting point, $y(0)=0=y(T)$. An integration over $x^\mu_0$ displaces a given path to every position in spacetime, which also makes translational invariance and as a consequence four-momentum conservation manifest. Then we get the effective action in the worldline formalism
\cite{Strassler:1992zr,Dietrich:2013kza}, 
\begin{align}
w
=
&\int \D^4x_0\int_{\varepsilon>0}^\infty\frac{\D T}{T^3}\;\E^{-m^2T}\;\Lag,
\label{eq:ads?}
\\
\Lag
=
\frac{\mathcal{N}}{(4\pi)^2}
&\int_\mathrm{P}[\D y]\;\E^{-\int_0^T\D\tau[\frac{\dot y^2}{4}+\I \dot y \cdot V(x_0+y)]} .
\label{eq:lag}
\end{align}
The effective action appears in the form of a Lagrangian density $\mathcal{L}$, which is integrated over an AdS$_5$ space that is soft-wall warped \cite{Karch:2006pv}. The Schwinger proper time $T$ groups with the four usual spacetime dimensions in an AdS$_5$ space with the metric parametrisation
\be
\D s^2\overset{g}{=}-\frac{\D T^2}{4T^2}+\frac{\D x\cdot\D x}{T}.
\label{eq:patch}
\ee
As already mentioned before, the isometries of AdS$_5$, SO(4,2), are the same as of the conformal group of 3+1-dimensional Minkowski space, which is inherited by massless classical gauge field theories. This way it is understandable, how this structure arises here, but it is still not trivial that the conformally invariant measure factors from the breaking (warping) contributions so as to become quite evident.
[Interestingly, a similar integration measure comes about when integrating over equivalence classes of (gauge) instanton configurations \cite{'tHooft:1976fv}, where $\D^4x_0$ represents the translations and $\frac{\D T}{T^3}$ for the dilatations (different powers of $T$ may appear depending on the details). Here, we find this kind of integration measure expressly without integrating over gauge field configurations. In exchange, we integrated out matter fields, which demonstrably leads to a very similar structure.]
The proper-time regularisation $T\ge\varepsilon>0$ reminds of the UV-brane regularisation in holography. The Lagrangian density $\mathcal{L}$ is made up of the path integral over all closed paths, $y^\mu(0)=y^\mu(T)$, on the interval $\tau\in[0;T]$. Thus, we are in a quantum mechanical setting like in other holographic approaches. The interaction part takes the form of a Wilson loop $\E^{-\I\oint\D y\cdot V}$. This makes the local invariance under the transformation $V^\mu\rightarrow\Omega[V^\mu+\I\Omega^\dagger(\partial^\mu\Omega)]\Omega^\dagger$ manifest, and hidden local symmetry \cite{Bando:1984ej} emerges.

In order to compute correlation functions of the source, we have to integrate out the position field $y^\mu$. To this end, we expand the interaction part in a Taylor series and Fourier transform the source $V^\mu$,
\begin{align}
&\E^{-\I\int_0^T\D\tau\:\dot y \cdot V(x_0+y)}
=\\=
{}&\sum_{n=0}^\infty\frac{(-\I)^n}{n!}\bigg[\int_0^T\D\tau\:\dot y \cdot V(x_0+y)\bigg]^n
=\\=
{}&\sum_{n=0}^\infty\frac{(-\I)^n}{n!}\bigg[\int_0^T\D\tau\int\frac{\D^4q}{(2\pi)^4}\E^{-\I q\cdot(x_0+y)}\:\dot y \cdot\tilde V(q)\bigg]^n .
\nonumber
\end{align}
The $n$th power of the integrals can be written as
\begin{align}
&\prod_{j=1}^n\int_0^T\D\tau_j\int\frac{\D^4q_j}{(2\pi)^4}\:(\dot y_j \cdot\tilde V_j)\:\E^{-\I\sum_{i=1}^n q_i\cdot(x_0+y_i)}
\nonumber\subset\\\subset
{}&\prod_{j=1}^n\int_0^T\D\tau_j\int\frac{\D^4q_j}{(2\pi)^4}\:\E^{\sum_{i=1}^n [\dot y_i \cdot\tilde V_i-\I q_i\cdot(x_0+y_i)]}.
\end{align}
Here $y_j=y(\tau_j)$, and ``$\subset$'' indicates that only terms linear in each of the $n$ different $\tilde V_j=\tilde V(q_j)$ contribute, where $\tilde V(q)$ is the Fourier transform of $V(x)$. With this rewrite the path integral is Gaussian, and we can carry it out by completing the square,
\begin{align}
\mathcal{N}\int_\mathrm{P}[\D y]\;\E^{\int_0^T\D\tau[-\frac{\dot y^2}{4}+\sum_{j=1}^n (\dot y_j \cdot\tilde V_j-\I q_j\cdot y_j)\delta(\tau-\tau_j)]}
=\\=
\E^{\frac{1}{2}\sum_{i,j=1}^n(G_{ij}q_i\cdot q_j+2\I\dot G_{ij}\tilde V_i\cdot q_j+\ddot G_{ij}\tilde V_i\cdot\tilde V_j)}.
\end{align}
Here
\be
G_{ij}=G(\tau_i,\tau_j)=G(\tau)=T\hat\tau(1-\hat\tau)
\ee
stands for the worldline propagator, where $\tau=|\tau_i-\tau_j|$ and $\hat\tau=\frac{\tau}{T}$. $\dot G_{ij}$ and $\ddot G_{ij}$ represent its first and, respectively, second derivative with respect to the first variable. $G$ satisfies the equation of motion
\be
\partial^2_{\tau_1}G(\tau_1,\tau_2)=2\delta(\tau_1-\tau_2)-\frac{2}{T}.
\label{eq:wlpeom}
\ee
The constant inhomogeneity $-\frac{2}{T}$ corresponds to a spread out counter charge, is needed to have a finite Poisson problem on a compact space, and is consistent with $\int_0^T\D\tau\,y^\mu=0$. We could also start out with an arbitrary constant counter charge. Then the discontinuity at $\tau_1=\tau_2$ together with periodicity and symmetry under the exchange $\tau_1\leftrightarrow\tau_2$ fix the constant uniquely. As yet another alternative, we do not have to resort to defining the worldline propagator through a differential equation at all, but can directly integrate over all periodic functions, which is done in (\ref{eq:Galt}).

Putting everything together the Lagrangian density is given by \cite{Schubert:2001he}
\begin{align}
\mathcal{L}
\subset
{}&\frac{1}{(4\pi)^2}\sum_{n=0}^\infty\frac{(-\I)^n}{n!}\int\frac{\D^{4n}q}{(2\pi)^{4n}}\int_0^T\D^n\tau\:\E^{-\I x_0\cdot\sum_{j=1}^n q_j}
\times\nonumber\\&\times
\E^{\frac{1}{2}\sum_{i,j=1}^n(G_{ij}q_i\cdot q_j+2\I\dot G_{ij}\tilde V_i\cdot q_j+\ddot G_{ij}\tilde V_i\cdot\tilde V_j)},
\label{eq:bk}
\end{align}
where 
$
\int\frac{\D^{4n}q}{(2\pi)^{4n}}\int_0^T\D^n\tau
=
\prod_{j=1}^n
(\int\frac{\D^4q_j}{(2\pi)^{4n}}\int_0^T\D\tau_j) .
$
For plane-wave sources this is the master formula of Bern and Kosower \cite{Bern:1991aq} for $n$-point correlation functions. 

The two-point function, for example, reads 
\begin{align}
w_2
&=
-\frac{1}{32\pi^2}\int\frac{\D^4q}{(2\pi)^4}\int^\infty_\varepsilon\frac{\D T}{T}\:\E^{-m^2T}\int_0^1\D\hat\tau_1\D\hat\tau_2\:
\times\nonumber\\&\times
(-q^2)\tilde\Pi^{\mu\nu}(q)\E^{-G_{12}q^2}\tilde V_\mu^*(q)\tilde V_\nu(q) \dot G_{12}^2 ,
\label{eq:s2}
\end{align}
where $\tilde\Pi^{\nu\lambda}(q)=\frac{q^\nu q^\lambda}{q^2}-\eta^{\nu\lambda}$, making transversality and thus local invariance manifest. (Up to a numeric prefactor the result for fermionic quarks is obtained by replacing $\dot G_{12}^2\rightarrow G_{12}$.) The dominant contribution for small $\varepsilon$ is given by
\be
\int_\varepsilon^\infty\frac{dT}{T}\E^{-T[m^2+q^2(\hat\tau-\hat\tau^2)]} 
\approx
-\gamma
-\ln\{\varepsilon[m^2+q^2(\hat\tau-\hat\tau^2)]\} .
\label{eq:forcomp}
\ee
As soon as $-q^2>4m^2$, the correlator develops an imaginary part. This signals that then $V^\mu$ can decay into its constituents. Thus, the worldline formulation lives naturally in a warped AdS$_5$ space and features hidden local symmetry, but in the absence of binding energy it does not show any spectrum of hadrons.

\subsection{AdS/QCD\label{sec:adsqcd}}

Let us compare this with the two-point function in soft-wall AdS/QCD. It is encoded in the quadratic action
\be
S^{5D}_2=-\frac{1}{4}\int\D^4x\frac{\D T}{2T^3}\E^{-\m^2T} g^{ab}g^{cd}\mathcal{V}_{ac}\mathcal{V}_{bd}\: ,
\label{eq:s5d}
\ee
which must be evaluated on the classical solution. Here $g^{\mu\kappa}$ stands for the inverse AdS$_5$ metric belonging to the parametrisation (\ref{eq:patch}), and $\m$ is the warp parameter. The sources in four dimensions are promoted to fields in five dimensions, $V_\lambda(x)\rightarrow\mathcal{V}_\lambda(x,T)$. As usual, we shall work in axial gauge $\mathcal{V}_5\equiv0$. On the classical solution there remains only a surface term at $T=\varepsilon>0$,
\be
\breve S^{5D}_2
=
\frac{1}{4}\int\frac{\D^4p}{(2\pi)^4} 
\tilde\Pi^{\nu\lambda}(p)\tilde V^*_\nu(p)\tilde V_\lambda(p)[\partial_T \tilde{\breve{v}}(p,\varepsilon)] ,
\label{eq:s5dsaddle}
\ee
where we have already imposed the boundary condition
\be
\tilde v(p,\varepsilon)=1,\mathrm{~where~}\tilde{\breve{\mathcal{V}}}_\lambda(p,T)=\tilde V_\lambda(p)\tilde v(p,T), 
\label{eq:inicond}
\ee
which links the sources in four dimensions to the fields in five.
Before we continue with the evaluation of the action on the saddle point, let us take a look at the effective action $w$ in the so-called inverse mass expansion \cite{Schubert:2001he}, which effectively is a Taylor expansion in $T$ of the Lagrangian (\ref{eq:lag}). To lowest nontrivial order,
\be
w_{I\!I}
=
\frac{-1}{6(4\pi)^2}
\int\D^4x_0\,\frac{\D T}{2T^3}\:\E^{-m^2T}g^{\mu\kappa}g^{\nu\lambda}V_{\mu\nu}V_{\kappa\lambda} ,
\label{eq:T2}
\ee
which we express with the help of the inverse AdS$_5$ metric $g^{\mu\kappa}$. (\ref{eq:T2}) coincides with (\ref{eq:s5d}) for $\tilde v(p,T)\equiv 1$ up to an overall normalisation, which can be absorbed into the normalisation of the fields or a coupling constant. This similarity gives another strong indication for the close relationship between the two approaches, and at the end of Sect.~\ref{sec:wilson} we will use it to exactly reproduce the AdS/QCD result, thus closing the circle.

Continuing with the evaluation of (\ref{eq:s5dsaddle}), 
instead of the second boundary condition one demands the solution be normalisable. The normalisable solution for the equation of motion 
\be
\Big(4\partial_T^2+\frac{p^2}{T}-\m^4\Big)\E^{-\m^2T/2}\tilde{\breve{\mathcal{V}}}^\perp(p,T)=0,
\label{eq:cleom}
\ee
is given by
\be
\tilde{\breve v}(q,T)
=
\Gamma\Big(1-\frac{q^2}{4\mu^2}\Big)\:T\:U\Big(1-\frac{q^2}{4\mu^2},2,\mu^2T\Big),
\ee
where the first boundary condition has already been taken into account and $U$ stands for Kummer's $U$ function. The second independent but non-normalisable solution would have contained Kummer's $M$ function. The derivative that enters the surface term reads
\be
\partial_T\tilde{\breve{v}}(p,\varepsilon)
=
-\frac{p^2}{4}[\ln(\varepsilon \m^2)+\gamma+\psi(1-\sfrac{p^2}{4\m^2})+\gamma] ,
\label{eq:dv}
\ee
where $\psi$ represents the digamma function. After the identifications $\m^2\leftrightarrow m^2$ and $-p^2\leftrightarrow q^2$ the behaviour for small $\varepsilon$ coincides with the above worldline result for $|p^2|\ll4\m^2$. Thus use of the same symbol $\varepsilon$ for both, the position of the UV brane and the proper-time regularisation, was justified, as they both parametrise the same kind of regularisation. The UV finite piece of (\ref{eq:dv}) can be expressed as
\be
\gamma+\psi(1-\sfrac{p^2}{4\m^2})
=
\m^2\int_0^\infty\D T\:\frac{\E^{-\m^2T}-\E^{-(\m^2-\frac{p^2}{4})T}}{1-\E^{-\m^2T}} ,
\label{eq:denominator}
\ee
with an obvious representation as geometric series, which bears testimony to the presence of a tower of states with equal spacing between the squared masses, $q^2=4nm^2$, $n\in\mathbbm{N}$. (See, e.g., Fig.~1 in \cite{Karch:2006pv}.) Each addend consists of a unilateral Laplace transforms. 

\section{Hadrons on the worldline\label{sec:hotwl}}

After the above preparations let us try to implement such an integration measure in (\ref{eq:denominator}) also in (\ref{eq:s2}) by means of a change of variables.
(Strictly speaking, one could argue that this amounts to adding a second phenomenological ingredient, i.e., the linearly spaced tower of states, but we will argue that the latter is also emergent \cite{Dietrich:2012un}, thus this apparent addition is only temporary.) The fitting substitution is given by
\be
\label{eq:covT}
cT=\E^{c\Theta}-1.
\ee
It leads to
\be\int_\varepsilon^\infty\frac{\D T}{T}f(T)
=
c\int_{\varepsilon}^\infty\D\Theta\frac{f[T(\Theta)]}{1-\E^{-c\Theta}}.
\label{eq:tower}
\ee
(In what follows we set $m=0$.) Also the upper bound of integration in the worldline action (\ref{eq:lag}) is affected. It can be undone---in the sense that after the substitution the upper bound is equal to the `extradimensional' variable $\Theta$, just as it was equal to $T$ before carrying out (\ref{eq:covT})---by a subsequent replacement
\be
c\tau=\E^{c\theta}-1 .
\label{eq:covtau}
\ee
The latter affects the integrand of the worldline action,
\be
\int_0^T\D\tau\Big(\frac{\D y}{\D\tau}\Big)^2
=
\int_0^\Theta\D\theta\:\E^{-c\theta}\Big(\frac{\D y}{\D\theta}\Big)^2 .
\ee
With the aim of restoring a standard kinetic term we also change the field variable,
\be
y^\mu=\E^{c\theta/2}\xi^\mu.
\ee
The result is
\be
\int_0^T\D\tau\Big(\frac{\D y}{\D\tau}\Big)^2
=
\int_0^\Theta\D\theta\Big[\Big(\frac{\D\xi}{\D\theta}\Big)^2+\frac{c^2}{4}\xi^2+c\cancelto{0}{\frac{\D(\xi^2)}{\D\theta}}\Big],
\label{eq:coco}
\ee
where the surface term, which arises from the total derivative in the last addend drops out when using $y^\mu(0)=0=y^\mu(T)$. A standard kinetic term has been restored but is now accompanied by a repulsive harmonic oscillator potential. The correlation functions for $V^\mu$ computed using these variables are of course identical to those computed using the original variables. After all, we have only carried out a change of variables. Consequently, we can conclude that the repulsive harmonic oscillator compensates exactly the effect from the tower of states, which we had introduced into the integration measure. 
As a consequence, to have a net physical effect one must use a detuned setup with different $c$s in the tower and in the harmonic oscillator, including the more minimal cases where one of the parameters is zero.

We can confirm this conclusion by relating the transformations (\ref{eq:covT}), (\ref{eq:covtau}), and (\ref{eq:coco}) to those carried out in \cite{de Alfaro:1976je},
\begin{align}
\D\Theta&=\frac{\D T}{U(T)} ,\\
\D\theta&=\frac{\D\tau}{U(\tau)} ,\\
\xi^\mu&=\frac{y^\mu}{\sqrt{U(\tau)}},\\
U(\tau)&=c_0+c_1\tau+c_2\tau^2 .
\label{eq:covdaff}
\end{align}
They coincide for the choice $c_0=1$, $c_1=c$, and $c_2=0$, and relate the above unilateral Laplace to the corresponding Mellin transformations. For general real values of $c_0$, $c_1$, and $c_2$ the potential can have both signs, $+\frac{1}{4}\xi^2(c_1^2-4c_0c_2)$, a fact that has been exploited in \cite{Brodsky:2013ar}. Interestingly this obviously introduces a scale into the Lagrangian density. The action, however, does not lose conformal invariance, as the time variable is adjusted accordingly \cite{de Alfaro:1976je}. This corroborates our above assessment of the precise cancellation of effects between our change of variables and the induced repulsive harmonic oscillator.

Also the interaction term is affected by the substitutions,
\be
\oint\D y\cdot V
=
\oint\D\xi\cdot\frac{\partial y}{\partial\xi}\cdot V ,
\ee
where $\frac{\partial y_\mu}{\partial\xi_\nu}=\delta_\mu^\nu\E^{c\theta/2}$. The extra factor can be absorbed in a redefinition of the source
\be
\int\D\tau\:\dot\xi\cdot V
=
\oint\D\xi\cdot\frac{\partial y}{\partial\xi}\cdot V
=
\oint\D\xi\cdot W
=
\int\D\theta\,\acute\xi\cdot W ,
\ee
where $\acute\xi=\frac{\partial\xi}{\partial\theta}$.
The source $W$ has the same value on the UV boundary as $V$, because $W\stackrel{\theta=0}{=}V$. This rescaling of the source is known from AdS/QCD soft-wall calculations, where it moves the warping away from the kinetic term \cite{Karch:2006pv,Dietrich:2008ni}.

\subsection{Relation to two-body interaction\label{sec:twobody}}

The harmonic oscillator in the worldline action corresponds to a two-body interaction,
\begin{align}
&\int_0^1\D\hat\tau_1\D\hat\tau_2\:[y(\tau_1)-y(\tau_2)]^2
=\\=
{}&\int_0^1\D\hat\tau_1\D\hat\tau_2\:\{[y(\tau_1)]^2+[y(\tau_2)]^2-2\:y(\tau_1)\cdot y(\tau_2)\}
\nonumber=\\=\nonumber
{}&1\times\int_0^1\D\hat\tau_1\:[y(\tau_1)]^2+1\times\int_0^1\D\hat\tau_2\:[y(\tau_2)]^2
-\\&-
2\cancelto{0}{\int_0^1\D\hat\tau_1\:y(\tau_1)}\cdot\cancelto{0}{\int_0^1\D\hat\tau_2\:y(\tau_2)}
=\\=
{}&2\int_0^1\D\hat\tau\, [y(\tau)]^2 ,
\label{eq:twobody}
\end{align}
where we used the centre-of-mass convention $\int_0^1\D\hat\tau\, y^\mu(\tau)=0$. 

A potential like in (\ref{eq:twobody}) is also found in another framework for describing hadrons \cite{Dietrich:2012un}: The equations of motion for the gauge field $A^\mu$, which necessarily accompanies any given charge configuration, in Coulomb gauge do not only admit the instantaneous Coulomb solution, but also an additional component of $A^0$ that is linear in the distance between the constituents (here of a meson for the sake of concreteness). Thus, to start with, it preserves translational invariance and, as a matter of fact, full Poincar\'e invariance, which only survives for the linear potential. Rotational invariance and stationarity of the action are assured if the linear potential is aligned with the constituents. The field energy due to the interference between the Coulomb and the linear component is finite if the state is charge neutral. The linear field's magnitude is set by a boundary condition, ${F_{\mu\nu}}^2\rightarrow\Lambda^2$ when $|\mathbf{x}_1-\mathbf{x}_2|\rightarrow\infty$ \cite{Dietrich:2012un}. 
(This and the omission of gauge corrections resembles the condensate framework of \cite{Shifman:1978bx}.) 
In quantum electrodynamics the effect of such a term is not observed. As a consequence, there $\Lambda$ must be tiny. In QCD where, after all, dimensional transmutation does take place it may well be present and even required to describe observed phenomena. Additionally, the linear component is the contribution with the lowest power in the coupling constant and in a setting, where perturbation theory can be applied it can be used to compute the hadronic `Born term' \cite{Dietrich:2012un}. (There are actually indications that in the IR the behaviour of the QCD coupling constant does not forestall a perturbative counting scheme \cite{Dokshitzer:1998qp}. Furthermore, the spectra of positronium and charmonium are qualitatively very similar despite showing vastly different energy scales. Last but not least, hadronic states can be characterised by their valence quark content, and we do not see a plethora of multiquark states, hybrids or glueballs.)

In order to incorporate the effect of this linear potential in the present framework, we note first that the gauge field $A^\mu$ would appear in (\ref{eq:lag}) exactly where the vector source $V^\mu$ sits. Thus, after replacing the source by the gauge field we integrate out the latter using the corresponding weight function,
\begin{align}
&\mathcal{N}_A\langle\E^{-\I\int_0^T\D\tau\,\dot y \cdot A}\rangle
=\\=
{}&\mathcal{N}_A\int[\D A]\,\E^{-\frac{1}{2}\int\D^4x\,A\cdot\Gamma^{-1}\cdot A}\E^{-\I\int_0^T\D\tau\,\dot y \cdot A}
=\\=
{}&\E^{-\frac{1}{2}\int_0^T\D\tau_1\D\tau_2\,\dot y_1\cdot\Gamma(y_1-y_2)\cdot\dot y_2}
\rightarrow\\\rightarrow
{}&\E^{\frac{\Lambda}{2}\int_0^T\D\tau_1\D\tau_2\,\delta(y_1^0-y_2^0)\dot y_1^0|\mathbf{y}_1-\mathbf{y}_2|\dot y_2^0}.
\label{eq:lin}
\end{align}
Here $\Gamma$ is the propagator of the gauge field, which in the hadronic `Born' approximation \cite{Dietrich:2012un} encodes the instantaneous linear potential [see (\ref{eq:lin})]. The normalisation $\mathcal{N}_A$ cancels the average over the case where $\dot y^\mu\equiv 0$, i.e., $\mathcal{N}_A\times\langle 1\rangle=1$. The above averaging corresponds to incorporating all exchanges of gauge bosons in diagrams like those shown in Fig.~\ref{fig}. To the contrary, neither quark loops (i.e., we are in the quenched case, which is a characteristic in common with \cite{'tHooft:1973jz}, where quark loops are subleading) nor gauge loops are included. As always, exclusively periodic trajectories contribute to the path integral. Then for paths that turn back in the time direction exactly once, i.e., for Fock states without additional pairs as in \cite{Dietrich:2012un}, we have
\begin{align}
\mathcal{N}_A\langle\E^{-\I\int_0^T\D\tau\,\dot y \cdot A}\rangle
\supset
\E^{\frac{\Lambda}{2}\int_0^T\D\tau\,\mathrm{sgn}(\dot{\bar{y}}^0)\dot y^0|\mathbf{y}-\bar{\mathbf{y}}|}
=\E^{-\Lambda\times\mathrm{Area}},
\label{eq:area}
\end{align}
where the exponent equals $\Lambda$ times the absolute surface area enclosed by the path. (Similarly, a constant external magnetic field gives a proportionality to the directed surface in this place. A constant magnetic field is known to entail linearly spaced Landau levels, which become evident in a formulation as harmonic oscillator in a composite variable \cite{Greiner:1992bv}. This has been commented on in \cite{Cornwall:2003mt}.)
Here $\bar{\mathbf{y}}=\mathbf{y}[\bar\tau(\tau)]$, i.e., the spatial coordinate where the time coordinates coincide, $\bar{y}^0=y^0[\bar\tau(\tau)]\overset{!}{=}y^0(\tau)$. We can complete the square to find
\begin{align}
&(\dot y^0)^2-2\Lambda\:\mathrm{sgn}(\dot{\bar{y}}^0)\dot y^0|\mathbf{y}-\bar{\mathbf{y}}|
=\\=
{}&(\dot z^0)^2-\Lambda^2(\mathbf{y}-\bar{\mathbf{y}})^2
=\\=
{}&(\dot z^0)^2-\Lambda^2({y}-\bar{{y}})^2 ,
\end{align}
where $\dot z^0=\dot y^0-\Lambda\,\mathrm{sgn}(\dot{\bar{y}}^0)|\mathbf{y}-\bar{\mathbf{y}}|$ and $\dot{\bar{y}}^0=\partial_\tau y^0|_{\tau=\bar\tau}$.
In the last step we made use of the fact that by definition $y^0-\bar{y}^0\equiv0$ and added a zero to emphasise the resemblance with (\ref{eq:twobody}). [It is always $\mathrm{sgn}(\dot{\bar{y}}^0)=-\mathrm{sgn}(\dot{{y}}^0)$. For parametrisations where also $\dot y^0=-\dot{\bar{y}}^0$---a natural gauge choice in an equal-time approach---$z^0$ is periodic and $z^0-\bar z^0\equiv0$ as well.]
Correspondingly, the bound states in this approach are spaced linearly \cite{Dietrich:2012un}, and, at that, without imposing postulate (a)!

Furthermore, $\dot z^0$ matches the zero component of the kinematical momentum $\Pi^\mu$ defined in \cite{Dietrich:2012un}, where it allows for introducing a Lorentz invariant evolution variable and thus to solve the bound state equation for all Lorentz frames at once. Here $\dot z^0$ equals the canonical momentum conjugate to $x^0$, obtained by taking the functional derivative of the worldline action with respect to $\dot x^0$, up to a numerical factor. 

\section{Wilson flow\label{sec:wilson}}

Let us take another look at the two-point function (\ref{eq:s2}). After an integration by parts,
\be
w_2
=
\frac{-1}{32\pi^2}\int\frac{\D^4q}{(2\pi)^4}\int^\infty_\varepsilon\frac{\D T}{T^2}\int_0^T\D\tau\:\ddot G\:\tilde V_\mu^*(q)\E^{-Gq^2}\tilde V^\mu_\perp(q) ,
\label{eq:w2}
\ee
where $\tilde V^\mu_\perp(q)=\tilde\Pi^{\mu\nu}(q)\tilde V_\nu(q)$. We can define $\tilde V^\mu_\perp(q,G)=\E^{-Gq^2}\tilde V^\mu_\perp(q)$, which solves
\be
(\partial_{G}+q^2)\tilde V^\mu_\perp(q,G)=0,
\ee
for the initial condition
\be
\tilde V^\mu_\perp(q,G=0)=\tilde V^\mu_\perp(q).
\label{eq:flowinicond}
\ee
For comparison in position space,
\be
V^\mu_\perp(x,G)=\E^{G\Box}V^\mu_\perp(x)=\int\frac{\D^4 x^\prime}{(4\pi G)^2}\,\E^{-\frac{(x-x^\prime)^2}{4G}}V^\mu_\perp(x^\prime).
\label{eq:position}
\ee
Hence $V(x,G)$ arises from the source $V(x,0)$ by Gaussian smearing. The smoothness (width of the Gaussian) is a function of the separation in the extra dimension. As a matter of fact, (\ref{eq:position}) is a solution of
\be
(\partial_{G}-\Box)V^\mu_\perp(x,G)
=
\partial_{G}V^\mu_\perp(x,G)-\partial_\nu V^{\nu\mu}(x,G)
=
0,
\label{eq:flowspace}
\ee
i.e., the defining differential equation of the Wilson flow (gradient flow) $V^\mu_\perp(x,G)$ \cite{Luscher:2009eq}, where $V^{\mu\nu}$ is the field strength constructed from $V^\mu$. In the Wilson flow the flow time also represents a fifth auxiliary variable. 
Here the fifth dimensional separation $\tau$ does not appear directly.
Instead the worldline propagator $G$ appears as flow-time interval. This is caused by the periodicity of the paths contributing to the effective action $w$, which leads to nonlinear terms in $G$: On the real line the equation of motion for the worldline propagator would simply be the one-dimensional Poisson equation $\partial_{\tau_1}^2G(\tau_1,\tau_2)\overset{\mathbbm{R}}{=}\delta(\tau_1-\tau_2)$ with its (piecewise) linear solution. This solution, however, definitely does not have period $T$. In order to have a periodic solution, we must include a counter charge \cite{Strassler:1992zr} (also already because we are dealing with a Poisson problem on a finite interval). Consequently the worldline propagator acquires nonlinear terms. That the worldline propagator  takes the place of the flow time implies that the flow, defined for any value of the flow time, is only sounded from $G(0)=0=G(T)$ to $G(T/2)=T/4$. (For $\tau>\frac{T}{2}$ the distance decreases again.) The integrations over $T$ and $\tau$ correspond to a linear superposition of the flow at various flow times. These results are not altered qualitatively when considering fermionic instead of scalar quarks, as the factor of $E^{-q^2G}$ in $w_2$ stays the same.

Alternatively, as defining equation for the flow we could also use
\be
(\partial_\tau+\dot G q^2)\tilde V^\mu_\perp(q,\tau)=0
\label{eq:diffq}
\ee
or equivalently
\be
(\partial_\tau-\dot G\Box)V^\mu_\perp(x,\tau)=0.
\label{eq:diffx}
\ee
Here the flow time is directly the distance $\tau$ in the fifth dimension, but the flow has a flow-time dependent `diffusion constant' $\dot G$. On the interval $\tau\in[0;T]$ the diffusion constant changes sign once. Noticing, however, that the integrand in (\ref{eq:w2}) is even under the interchange $\tau\leftrightarrow T-\tau$ we can take two times the integral over $\tau\in[0;\frac{T}{2}]$ on which $\dot G$ does not change sign, but only freezes in close to $\frac{T}{2}$.

For the sake of symmetry and for preparing the ground for a later generalisation to other correlators we can introduce the split $\E^{-q^2G}=\int_0^1\D\hta\E^{-q^2\hta G}\E^{-q^2(1-\hta)G}$, such that
\be
\tilde V_\mu^*(q)\E^{-Gq^2}\tilde V^\mu_\perp(q)
=
\int_0^1\D\hta\,\tilde V_\mu^*(q,\hta G)\tilde V^\mu_\perp[q,(1-\hta)G] .
\ee
(Additionally we could in fact plug in any function over $\check\tau\in[0;1]$ which is normalised to 1.)

Including, for example, an attractive harmonic oscillator,
\be
\mathcal{L}
\rightarrow
\frac{\mathcal{N}_c}{(4\pi)^2}
\int_\mathrm{P}[\D y]~\E^{-\int_0^T\D\tau(\frac{\dot y^2}{4}-\frac{c^2}{16}y^2+\I \dot y\cdot V)} ,
\label{eq:wlho}
\ee
where the normalisation is defined by $\mathcal{N}_c\times\int_\mathrm{P}[\D y]~\E^{-\frac{1}{4}\int_0^T\D\tau(\dot y^2-\frac{c^2}{4}y^2)}=1$, the worldline propagator is given by 
\be
\frac{c}{2} H(\tau)=\sin({\textstyle\frac{c}{2}}|\tau|)-\sin({\textstyle\frac{c}{2}} T)\frac{1-\cos(\frac{c}{2}\tau)}{1-\cos(\frac{c}{2} T)}.
\label{eq:wlprop}
\ee
Its equation of motion reads
\be
\Big(\partial_{\tau_1}^2+\frac{c^2}{4}\Big)H(\tau_1,\tau_2)=2\delta(\tau_1-\tau_2)-\#,
\ee
where $\#$ is once again a counter charge, which must still be determined. The space of homogeneous solutions is spanned by $\E^{\pm\I\frac{c}{2}(\tau_1-\tau_2)}$. The constant inhomogeneity can be caught by an additive constant particular solution $-\frac{4\#}{c^2}$. $H$ must be symmetric under the interchange $\tau_1\leftrightarrow\tau_2$ and periodic under $\tau_j\rightarrow\tau_j+z_jT$, $z_j\in\mathbbm{Z}$, $j\in\{1;2\}$. Finally, the $\delta$ inhomogeneity implies that, while $H=0$ at $\tau_1=\tau_2$, its first derivative jumps by 2 at this point. All these conditions can be satisfied simultaneously, fix the three free parameters (two for the homogeneous solution and one for the counter charge) uniquely, and yield (\ref{eq:wlprop}). With the above requirements $H$ only depends on the combination $\tau_1-\tau_2$, which reflects reparametrisation invariance on the worldline. The value of the counter charge is given by
\be
\#=\frac{c}{2}\frac{1+\cos(\frac{c}{2}T)}{\sin(\frac{c}{2}T)}.
\ee
In the limit $c\rightarrow 0$ it goes smoothly to $\frac{2}{T}$, as it should to have the original potential free equation of motion (\ref{eq:wlpeom}).
Likewise, $H$ goes smoothly to $G$ in the limit $c\rightarrow 0$. 

\begin{figure}[t]
\centerline{%
\includegraphics[width=\columnwidth]{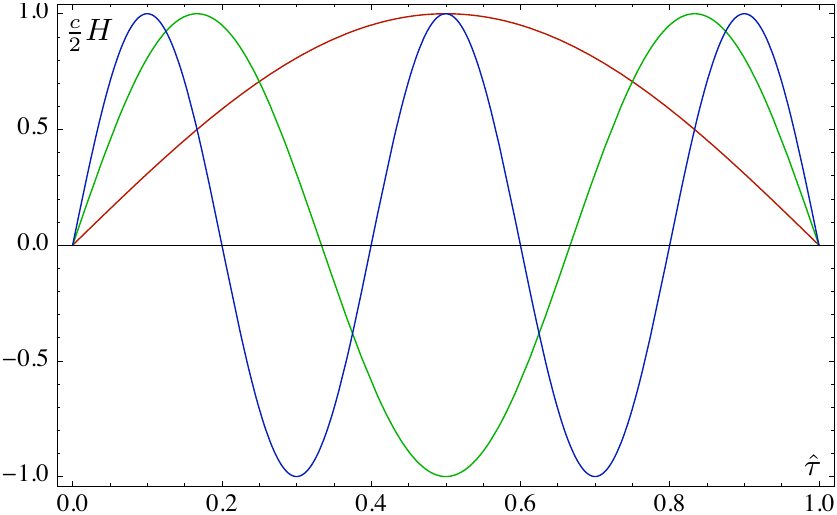}}
\caption{
$\frac{c}{2}H$ as a function of $\hat\tau$ with the number of nodes growing with $cT=2\pi$, $6\pi$, and $10\pi$, respectively. The initial slope equals $\frac{cT}{2}$. $\frac{c}{2}H$ for $cT=2\pi$ resembles $\frac{4}{T}G$. The identical amplitudes are due to the special values of $cT$.
}
\label{oscil}
\end{figure}

\begin{figure}[t]
\centerline{%
\includegraphics[width=\columnwidth]{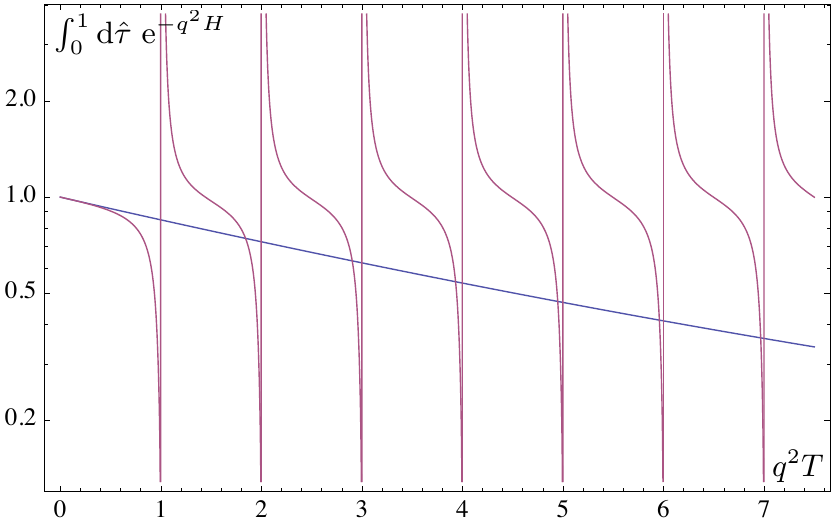}}
\caption{
$\int_0^1\D\hat\tau\;\E^{-q^2H}$ for $c=4\pi q^2$ (discontinuous graph)
and
$\int_0^1\D\hat\tau\;\E^{-q^2G}$ (continuous graph)
as functions of
$q^2T$.
}
\label{props}
\end{figure}

The repulsive case is obtained by the replacement $c\rightarrow\I c$. Then the worldline propagator becomes
\be
\frac{c}{2} h(\tau)=\sinh({\textstyle\frac{c}{2}}|\tau|)-\sinh({\textstyle\frac{c}{2}} T)\frac{1-\cosh(\frac{c}{2}\tau)}{1-\cosh(\frac{c}{2} T)}.
\label{eq:wlprop_rep}
\ee
Accordingly,
\be
\#\rightarrow\frac{c}{2}\frac{1+\cosh(\frac{c}{2}T)}{\sinh(\frac{c}{2}T)}.
\ee
Qualitatively, $h$ is similar to $G$ inasmuch as both always grow from 0 at $\tau=0$ to a maximum at $\tau=\frac{T}{2}$ and symmetrically to this point decrease back to 0 at $\tau=T$. Quantitatively, the overall magnitude of $h$ grows less fast with $T$ than that of $G$. For comparison, $G(\frac{T}{2})=\frac{T}{4}$ and $h(\frac{T}{2})=\frac{2}{c}\tanh(\frac{cT}{16})=\frac{T}{4}\,\{1-O[(cT)^2]\}$. For large values of $cT$ the growth of $h$ levels out at $h(\frac{T}{2})<\frac{c}{2}$, while the maximum of $G$ keeps growing.
At variance with $G$, $H$ does not increase monotonously without bounds for growing $T$. ($h$ increases monotonously, but is bounded.)
When plugged into $\E^{-q^2G}$ the growth of $G$ leads to an exponential suppression at large $T$, while for $h$ also this factor levels out. 
For small values of $T$, $H$ increases faster with $T$ than $G$. Then, however, $H$ starts to oscillate (see Fig.~\ref{oscil}). 
To the contrary, poles are appearing in $\int_0^1\D\hat\tau\;\E^{-q^2H}$ for $cT$ equal to integer multiples of $4\pi$ (see Fig.~\ref{props}).

We obtain the flow equation and two-point function for the harmonic-oscillator case by replacing $G$ by $H$ or $h$. This determines the distribution of flow times, which contributes to the superposition of the composite-field Wilson flow that extends the four-dimensional sources into the fifth dimension. On one hand, at the poles of $H$, the flow extends infinitely far in the fifth dimension; on the other, the convergent behaviour of $h$ leads to a continuum of contributions that are not exponentially suppressed like for $G$.

In the interpretation with a flow-time dependent diffusion constant $\dot G$ according to (\ref{eq:diffq}) and (\ref{eq:diffx}), $\dot G=1-2\hat\tau$ is $T$ independent. The superposition of flow times is only due to the integration over the variable $\hat\tau$. $\dot h=\sinh[\frac{cT}{4}(1-2\hat\tau)]/\sinh(\frac{cT}{4})$ goes to zero exponentially, $\propto E^{-\frac{c\tau}{2}}$ for large $T$. $\dot H$ first increases from small to large values of $T$. Then, however, it will eventually change sign and start to oscillate even on the reduced interval $\tau\in[0;\frac{T}{2}]$.

\subsection{Exact coincidence with AdS/QCD\label{sec:repro}}

Taking stock, above we have seen that the holographic AdS$_5$ space emerges straight away in the worldline approach and that the soft-wall warping is linked to harmonic oscillator interactions in the worldline action. The extension of the four-dimensional fields into the fifth dimension proceeds via a Wilson flow (gradient flow). This flow depends on the interaction potential insofar as the weight for the distribution of depths at which it is probed changes with the worldline propagator, which depends on the interaction term. Consequently, every interaction term corresponds to a different superposition of flow.

We can also ask a slightly different question: Given a particular way of breaking the conformal symmetry, for instance, by a warp factor, what is the dominant contribution to the flow? To this end, we take the effective action (\ref{eq:lag}) and replace the source $V^\mu(x)$ by the flow $V^\mu(x,\tau)$. Subsequently, we look for the stationary point of the action where $\frac{\delta w}{\delta V(x,T)}\overset{!}{=}0$ and impose the usual initial condition that at $T=\varepsilon$ the flow takes the value of the four-dimensional source. Solving the full saddle point equation is a formidable task as almost always, but, like in AdS/QCD [see, for example, (\ref{eq:s5d})], let us consider a truncation. With the replacement $V^\mu(x)\rightarrow V^\mu(x,\tau)$ (\ref{eq:T2}) turns into (\ref{eq:s5d}) if we identify $V^\mu(x,T)\leftrightarrow\mathcal{V}^\mu(x,T)$. Then the variation with respect to the flow field yields the equation of motion (\ref{eq:cleom}), and the flow initial condition (\ref{eq:flowinicond}) implies that the flow action [(\ref{eq:lag}) with the replacement $V^\mu(x)\rightarrow V^\mu(x,\tau)$] on the saddle point coincides with (\ref{eq:s5dsaddle}). The normalisable solutions are selected by only allowing a nondiverging flow. 
This coincidence is remarkable, as calculating the optimal flow is identical to the AdS/QCD computation. It, however, necessitates a reinterpretation of the deformation parameter $m^2$. In the original computation, $m^2$ was the bare mass and led to a threshold behaviour as soon as $q^2>4m^2$. Now, as the equations obtained from varying the Wilson flow coincide with those  in the AdS/QCD computation, we know that the former approach yields a linearly spaced tower of states. Hence, the parameter is linked to the intrinsic dynamics of the system and is not an external mass. Since we are allowing for a self-consistent $T$ profile, the breaking becomes spontaneous in nature. Also, consistent with the above, from the phenomenological point of view the magnitude of $m^2$ in the computation with the flow should be in the range of the constituent mass \cite{Dietrich:2013kza}.

\section{Worldline instantons\label{sec:instanton}}

Before we finish, let us mention that we can also interpret (\ref{eq:wlho}) with the concept of worldline instantons \cite{Affleck:1982,Dunne:2005sx}. A worldline instanton is a periodic classical solution $\breve y^\mu$ of the worldline action,
\be
\left.\frac{\delta S}{\delta y^\mu}\right|_{\breve y^\mu}\overset{!}{=}0 .
\ee
This stationary phase gives a dominant contribution to the effective action.
Subsequently, fluctuations around these solutions are taken into account,
\be
S[y^\mu=\breve y^\mu+\dy^\mu]=S[\breve y^\mu]+\frac{1}{2}\left.\frac{\delta^2 S}{\delta y_\mu\delta y_\nu}\right|_{\breve y^\mu}\dy^\mu\dy^\nu+O(\dy^3),
\ee
\newpage\noindent
which makes this a semiclassical approximation. The classical equations of motion for $V^\mu\equiv0$,
\be
\ddot{\breve y}^\mu+\frac{c^2}{4}\breve y^\mu=0 ,
\label{eq:cleomho}
\ee
have periodic solutions only for discrete values $T=\frac{4\pi n}{c}$, $n\in\mathbbm{N}$, of Schwinger's proper time. For $V^\mu\equiv 0$ the action is quadratic and we can perform the path integral directly for all values of $T$. (This remains true if we treat the source perturbatively.) To this end we can parametrise the periodic worldline trajectory according to
\be
y^\mu
=
\sum_{n=-\infty}^{+\infty}a_n^\mu\;\E^{\I n\frac{2\pi}{T}\tau} ,
\ee
where we put $a_0^\mu=0$ to have $\int_0^T\D\tau\: y^\mu(\tau)=0$ [for $a_0^\mu\neq0$ see the appendix] and $a_{-n}^\mu=(a_n^\mu)^*$ to have a real $y^\mu$. Let us take a look at the two-point function, as it contains the worldline propagator, from which we can also construct all the higher correlators (see Sect.~\ref{sec:wlf}). Thus we plug the above parametrisation into the worldline action and add two Fourier phases. Thereafter we carry out the proper-time derivatives, integrate over the proper time $T$ (which corresponds to the integrations for a Fourier series), use the Kronecker $\delta$ to perform one summation, split the summation at 0, and complete the square,
\begin{widetext}
\begin{align}\label{eq:source}
&-\frac{1}{4}\int_0^T\D\tau\;\Big(\dot y^2-\frac{c^2}{4}y^2\Big)-\I[q_1\cdot y(\tau_1)+q_2\cdot y(\tau_2)]
=\\={}&
-\frac{1}{4}\sum_{n,n^\prime=-\infty}^\infty a_n\cdot a_{n^\prime}\bigg[-nn^\prime\Big(\frac{2\pi}{T}\Big)^2-\frac{c^2}{4}\bigg]\int_0^T\D\tau\;\E^{\I(n+n^\prime)\frac{2\pi}{T}\tau}
-
\I\sum_{n=-\infty}^{\infty}a_n\cdot(q_1\E^{\I n\frac{2\pi}{T}\tau_1}+q_2\E^{\I n\frac{2\pi}{T}\tau_2})
=\\={}&
-\frac{T}{4}\sum_{n,n^\prime=-\infty}^\infty a_n\cdot a_{n^\prime}\bigg[-nn^\prime\Big(\frac{2\pi}{T}\Big)^2-\frac{c^2}{4}\bigg]\delta_{n,-n^\prime}
-
\I\sum_{n=-\infty}^{\infty}a_n\cdot(q_1\E^{\I n\frac{2\pi}{T}\tau_1}+q_2\E^{\I n\frac{2\pi}{T}\tau_2})
=\\={}&
-\sum_{n=-\infty}^\infty\bigg\{|a_n|^2\frac{T}{4}\bigg[n^2\Big(\frac{2\pi}{T}\Big)^2-\frac{c^2}{4}\bigg]
+
\I a_n\cdot(q_1\E^{\I n\frac{2\pi}{T}\tau_1}+q_2\E^{\I n\frac{2\pi}{T}\tau_2})\bigg\}
=\\={}&
-\sum_{n=1}^\infty\bigg\{|a_n|^2\frac{T}{2}\bigg[n^2\Big(\frac{2\pi}{T}\Big)^2-\frac{c^2}{4}\bigg]
+
\I a_n\cdot(q_1\E^{\I n\frac{2\pi}{T}\tau_1}+q_2\E^{\I n\frac{2\pi}{T}\tau_2})
+
\I a_{-n}\cdot(q_1\E^{-\I n\frac{2\pi}{T}\tau_1}+q_2\E^{-\I n\frac{2\pi}{T}\tau_2})\bigg\}
\label{eq:source_}=\\={}&\label{eq:source__}
-\sum_{n=1}^\infty\bigg\{\frac{T}{2}\bigg[n^2\Big(\frac{2\pi}{T}\Big)^2-\frac{c^2}{4}\bigg]
\bigg(a_n+\I\frac{q_1\E^{-\I n\frac{2\pi}{T}\tau_1}+q_2\E^{-\I n\frac{2\pi}{T}\tau_2}}{\frac{T}{2}[n^2(\frac{2\pi}{T})^2-\frac{c^2}{4}]}\bigg)
\cdot
\bigg(n\rightarrow-n\bigg)
+
\frac{|q_1\E^{\I n\frac{2\pi}{T}\tau_1}+q_2\E^{\I n\frac{2\pi}{T}\tau_2}|^2}{\frac{T}{2}[n^2(\frac{2\pi}{T})^2-\frac{c^2}{4}]}\bigg\} .
\end{align}
\end{widetext}
In the present parametrisation the path integral reads 
\be
\int_\mathrm{P}[\D y]
=
\prod_{n=1}^\infty\int \D a_n^\mu\: \D (a_n^\mu)^*
=
\int [\D a] [\D a^*] .
\ee
Thus, carrying out the path integral leads to
\begin{align}
\mathcal{N}_c\int_\mathrm{P}[\D y]\E^{-\frac{1}{4}\int_0^T\D\tau\;(\dot y^2-\frac{c^2}{4}y^2)-\I(q_1\cdot y_1+q_2\cdot y_2)}
=\nonumber\\={}
\E^{-\sum_{n=1}^\infty\frac{2}{T}
\frac{q_1^2+q_2^2+2q_1\cdot q_2\cos[n\frac{2\pi}{T}(\tau_1-\tau_2)]}{n^2(\frac{2\pi}{T})^2-\frac{c^2}{4}}},
\label{eq:wlprop2}
\end{align}
where the normalisation $\mathcal{N}_c$ cancels the path integral for $q_1=0=q_2$. The exponent turns out to be a Fourier-series representation of the worldline propagator (\ref{eq:wlprop}),
\be
\frac{1}{T}\int_{0}^{T}\D\tau\:H(\tau)\cos\Big(n\frac{2\pi}{T}\tau\Big)
=
-\frac{2}{T}\frac{1}{n^2\frac{(2\pi)^2}{T^2}-\frac{c^2}{4}}.
\ee
Imposing momentum conservation $q_1+q_2=0$ in (\ref{eq:wlprop2}) the constant term in the exponent cancels the harmonic term at $\tau_1=\tau_2$. Hence, the initial condition $H(0)=0$ is satisfied.  
The aforementioned exactly periodic solutions of the saddle point equations (\ref{eq:cleomho}) coincide with the poles of the propagator. 
For negative $c^2$ the derivation in this section goes through in the same way to reproduce (\ref{eq:wlprop_rep}). There are no poles in this propagator and, consistently, also no periodic classical solutions.

It is instructive to also track the appearance of the worldline propagator $G$ in the $c\rightarrow 0$ case. There we have
\begin{align}
&
{-\frac{Tq^2}{\pi^2}\sum_{n=1}^\infty
\frac{1-\cos[n\frac{2\pi}{T}(\tau_1-\tau_2)]}{n^2}}
=\\={}&
{-\frac{Tq^2}{\pi^2}
\Big[\frac{\pi^2}{6}-\Big(\frac{\pi^2}{6}-\frac{2\pi^2\frac{\tau_1-\tau_2}{T}}{2}+\frac{(2\pi\frac{\tau_1-\tau_2}{T})^2}{4}\Big)\Big]}
\nonumber
=\\={}&
{-Tq^2
\Big[\frac{\tau_1-\tau_2}{T}-\Big(\frac{\tau_1-\tau_2}{T}\Big)^2\Big]},
\label{eq:Galt}
\end{align}
where we made use of (24.12.8) from \cite{as},
\be
\sum_{n=1}^\infty\frac{\cos(n\theta)}{n^2}=\frac{\pi^2}{6}-\frac{\pi\theta}{2}+\frac{\theta^2}{4}.
\ee

Interestingly, the worldline instanton approach can be related \cite{Dietrich:2007vw} to the Gutzwiller trace formula \cite{Gutzwiller:1971fy}. It characterises quantum mechanical systems (in general approximately, but exactly for quadratic actions) through classical objects, i.e., periodic orbits, stability matrices, Morse indices, in analogy to the classical representation of quantum field theory in the framework of holography.

\section{Summary\label{sec:summary}}

Holographic approaches appear to be a good approximation to the properties of hadrons \cite{Karch:2006pv,Da Rold:2005zs}. Here we have contributed to understanding why this could be more than a coincidence. We did not take any detour via gravity or string theory, but stayed within quantum field theory and started directly in the worldline formalism. We have selected the dominant contributions as those in which additional internal glue is suppressed, in accordance with experiment \cite{White:1994tj,Gunion:1972qi,Landshoff:1974ew,Okubo:1963fa,Dokshitzer:1998qp}. When introducing sources as precursors for hadrons, our effective action takes immediately the form of a Lagrangian density integrated over an AdS$_5$ spacetime that is warped by all contributions that break conformal symmetry. This is a guise the effective action takes in holographic descriptions as well. The fifth coordinate is the Schwinger proper-time interval for the affine parameter for the contributing particle trajectories. Hidden local symmetry is emergent. The proper-time regularisation corresponds to the UV-brane regularisation in AdS/QCD. 

Still, consistently, in the absence of binding energy the sources simply fall apart at the threshold. Therefore, we continued with a little experiment. We mimicked the linear tower of states present in soft-wall AdS/QCD by a change of variables. When propagating the effect of this substitution through the entire expression for the effective action, we end up with a repulsive harmonic oscillator potential. Then again a change of variable must not affect the physical results. Hence, we have concluded that the effect of the tower of states is compensated exactly by the repulsive harmonic oscillator. This interpretation is corroborated by the fact that for the worldline action our substitutions fall into a class of transformations that leaves conformal symmetry intact, despite the fact that it introduces a scale into the Lagrangian density \cite{de Alfaro:1976je}. At the level of the action the change in the `time' variable compensates for the presence of the scale. 

As a consequence, in order to have a net effect, the above compensating setup must be detuned. A model for hadrons can, for example, have only the tower of states in the measure or only a harmonic oscillator potential in the worldline action. At the latter form we can also arrive from a different direction: Against the same phenomenological background, in \cite{Dietrich:2012un} we had identified a `Born term' for hadrons, which essentially consisted of a linear interquark potential. There it resulted in a linearly spaced bound-state spectrum. Here its incorporation into the present formalism again led to a harmonic oscillator potential in the worldline action.

In holography the four-dimensional sources are extended to fields in five dimensions. In the current context this happens through a particular kind of Wilson flow (gradient flow) for the composite fields. The presence of the harmonic oscillator influences the weight for the superposition of flows at different flow times. (An alternative interpretation with a variable diffusion parameter is available as well.) Remarkably, adding a variational principle for the flow, reproduces exactly the AdS/QCD computation.

\section{Outlook}

We can go beyond the present computation systematically by incorporating dynamical gluons order by order in the framework of higher-loop effective actions.
The analysis can be generalised straightforwardly to different sources, like other mesons $\bar\psi\Gamma\psi$, but also baryons, the dilaton $\lambda F_{\mu\nu}F^{\mu\nu}$ or the axion $a F_{\mu\nu}\tilde F^{\mu\nu}$.
It also remains to compute and compare higher correlators and to discuss different gauge and flavour symmetries.
All that will be treated in later work.

The emergence of similar curved extradimensional formulations in the framework of the worldline formalism extends to many situations also outside relativistic quantum field theory. Again the form of the enlarged spacetime depends on the isometries of the physical system. The conformal Galilean symmetry of the Schr\"odinger equation in 3 spatial dimensions, for example, can be constructed by constraining (\ref{eq:lag}) without the term $\E^{-m^2 T}$ in 4+1 dimensional Minkowski space by imposing $p^+=m$ \cite{Son:2008ye}. Accordingly, $x^+$ plays the role of time. This results in a six-dimensional volume element (two extra dimensions!), which is $\propto T^{-7/2}$. 
This  is the volume element for the correct metric \cite{Son:2008ye,Balasubramanian:2008dm}. In the present conventions,
\be
\D^2s\overset{\mathrm{g}}{=}
-\frac{\D T^2}{4T^2}
+\frac{2(\D x^+)^2}{T^2}
+\frac{2\D x^+\D x^--\D\mathbf{x}\cdot\D\mathbf{x}}{T} ,
\ee
where $x^\pm=\frac{x^0\pm x^4}{\sqrt{2}}$ and $x^4$ is the coordinate of the other extra dimension. (The power of $T$ in the denominator of the $(\D x^+)^2$ term can be different \cite{Balasubramanian:2008dm} without influencing the volume element.)
In order to understand this let us recall how this volume element came about in the relativistic case. There, in addition to the factor of $\frac{1}{T}$ from the representation of the logarithm (\ref{eq:ln}), it was due to the mismatch in numbers of position and momentum integrals in the path integral (\ref{eq:pathint}), 
\be
\frac{1}{T}\int\D^4p\,\E^{-Tp^2}\propto T^{-3}\propto\sqrt{g}.
\ee
For the present non-relativistic case, we have similarly,
\be
\frac{1}{T}\int\D^3\mathbf{p}\,\D p^-\,\E^{-T(2mp^-+\mathbf{p}^2)}\propto T^{-7/2}\propto\sqrt{\mathrm{g}}.
\ee

Finally, for a superconductor introducing a potential would bind the electrons into Cooper pairs. In this context a tower of states has not been observed, as the 'partonic' excitation, i.e., the free electrons, has too low an energy. The composite vectors, i.e., the spin-1 Cooper pairs are present \cite{TripletSupercon}, though.

\section*{Acknowledgments}

The author would like to thank
Stan Brodsky,
Guy de T\'eramond,
Luigi Del Debbio,
Gia Dvali,
C\'esar Gomez,
Stefan Hofmann,
Paul Hoyer,
Michael Kopp,
Matti J\"arvinen,
Joachim Reinhardt,
Tehseen Rug,
Andreas Sch\"afer,
Christian Weiss, and
Roman Zwicky 
for inspiring and informative discussions.
The work of the author was supported by the Humboldt foundation.

\appendix

\section{$a_0^\mu\neq0$}

In the source term of (\ref{eq:source}) $a_0^\mu$ appears in $a_0\cdot(q_1+q_2)$, which vanishes after momentum conservation is enforced by integrating out $x_0^\mu$. This persists for any number of external momenta, i.e., for arbitrarily high correlators. Furthermore, $a_0^\mu$ drops out of the kinetic term, because it belongs to the constant mode. $a_0^\mu$ contributes, however, to the potential term. For the starting-point conventions $y^\mu(0)=0=y^\mu(T)$ we have $a_0^\mu={-}\sum_{n=1}^\infty(a_n^\mu+{a_n^\mu}^*)$. We can express (\ref{eq:source_}) and (\ref{eq:source__}) as
\begin{align}
&a^\top Ma - \I a^\top j
=\\={}&\nonumber
(a-\sfrac{\I}{2} M^{-1}j)^\top M(a-\sfrac{\I}{2} M^{-1}j)
+
\sfrac{1}{4}j^\top M^{-1}j
\end{align}
where $a$ is the vector of the coefficients $a_n$, with $a_n$ and ${a_n}^*$ grouped side by side. $M$ can be cast in the form 
\be
M=A+\frac{c^2T}{16}vv^\top,
\ee
where $v$ is a vector of ones, and $A$ is block diagonal with $2\times2$ submatrices
\be
A_n={\frac{T}{4}}\bigg[\frac{c^2}{4}-n^2\Big(\frac{2\pi}{T}\Big)^2\bigg]\left(\begin{array}{cc}0&1\\1&0\end{array}\right)
\ee
on the diagonal. Then \cite{Sherman:1949}
\be
M^{-1}
=
A^{-1}
-
\frac{\frac{c^2T}{16}A^{-1}vv^\top\mathrm{A}^{-1}}{1+\frac{c^2T}{16}v^\top A^{-1}v}.
\label{eq:shermor}
\ee
From 
\be
{\mathrm{A}_n}^{-1}={\frac{4}{T}}\bigg[\frac{c^2}{4}-n^2\Big(\frac{2\pi}{T}\Big)^2\bigg]^{-1}\left(\begin{array}{cc}0&1\\1&0\end{array}\right),
\ee
follows
\be
v^\top A^{-1}v
=
\frac{8}{T}\sum_{n=1}^\infty\bigg[\frac{c^2}{4}-n^2\Big(\frac{2\pi}{T}\Big)^2\bigg]^{-1}
\ee
such that
\be
1+\frac{c^2T}{16}v^\top A^{-1}v
=
\frac{cT}{4}\cot\Big(\frac{cT}{4}\Big).
\ee
The numerator in (\ref{eq:shermor}) decomposes into $2\times2$ matrices,
\begin{align}
&(A^{-1}vv^\top A^{-1})_{nn^\prime}
=\\=&{}\nonumber
\Big(\frac{4}{T}\Big)^2
\bigg[\frac{c^2}{4}-n^2\Big(\frac{2\pi}{T}\Big)^2\bigg]^{-1}
\bigg[\frac{c^2}{4}-(n^\prime)^2\Big(\frac{2\pi}{T}\Big)^2\bigg]^{-1}
\left(\begin{array}{cc}1&1\\1&1\end{array}\right)
\end{align}
Finally,
\begin{align}
\nonumber
&\sfrac{1}{4}j^\top M^{-1}j
=\\={}&
\frac{4q^2}{T}\sum_{n=1}^\infty\frac{1-\cos[n\frac{2\pi}{T}(\tau_1-\tau_2)]}{\frac{c^2}{4}-n^2(\frac{2\pi}{T})^2}
-\\&-
\frac{4q^2c}{T^2}\tan\Big(\frac{cT}{4}\Big)
\bigg[\sum_{n=1}^\infty\frac{\cos(n\frac{2\pi}{T}\tau_1)-\cos(n\frac{2\pi}{T}\tau_2)}{\frac{c^2}{4}-n^2(\frac{2\pi}{T})^2}\bigg]^2,
\nonumber
\end{align}
where we made use of momentum conservation $q_1=-q_2=q$.

\end{document}